\documentclass[11pt]{article}
\usepackage{moriond,epsfig}

\def\be{\begin{equation}}
\def\ee{\end{equation}}
\def\bea{\begin{eqnarray}}
\def\eea{\end{eqnarray}}

\newcommand{\jprlBase}  [1]     {Phys.\ Rev.\ Lett. \xspace}
\newcommand{\jprl}      [1]    {\jprlBase\ {\bf #1}}
\newcommand{\jprBase}        {Phys.\ Rev.\ \xspace}
\newcommand{\jprd}      [1]  {\jprBase\ D~{\bf #1}}
\newcommand{\plBase}   [1]         {Phys.\ Lett.\xspace}
\newcommand{\plb}      [1]    {\plBase\ B~{\bf #1}}
\newcommand{\nimBaseA}       {Nucl.\ Instr.\ Meth.\ \xspace }

\newcommand{\zpBase}         {Z.\ Phys.}

\newcommand{\npBase}         {Nucl.\ Phys.\ \xspace}

\newcommand{\progtp}    [1]  {{Prog.\ Theor.\ Phys.\ {\bf #1}}}
\newcommand{\jmplBase}  [1]     {Mod.\ Phys.\ Lett.}

\newcommand{\epjBase}  [1]     {Eur.\ Phys.\ J. \xspace}

\newcommand\etal{{\it et al.}}
\def\babar{{\em B}{\footnotesize\em A}{\em B}{\footnotesize\em AR}}

\newcommand{\mevcc}{\mbox{$\textrm{MeV}/c^2$}}

\newcommand{\mev}{\mbox{$\textrm{MeV}$}}

\def\CP                {\ensuremath{C\!P}\xspace}

\def\to                 {\ensuremath{\rightarrow}\xspace}
\newcommand{\calB}{\mbox{${\cal B}$}}
\newcommand{\calA}{\mbox{${\cal A}$}}
\def\BB{\mbox{$B\overline B\ $}}

\def\Dbar    {\kern 0.2em\overline{\kern -0.2em D}{}\xspace}

\def\Dz      {\ensuremath{D^0}\xspace}
\def\Dzb     {\ensuremath{\Dbar^0}\xspace}
\def\DzDzb   {\ensuremath{\Dz {\kern -0.16em \Dzb}}\xspace}
\def\Dp      {\ensuremath{D^+}\xspace}
\def\Dm      {\ensuremath{D^-}\xspace}

\def\DpDm    {\ensuremath{\Dp {\kern -0.16em \Dm}}\xspace}
\def\Dstar   {\ensuremath{D^*}\xspace}

\def\Dstarz  {\ensuremath{D^{*0}}\xspace}

\def\Dstarp  {\ensuremath{D^{*+}}\xspace}
\def\Dstarm  {\ensuremath{D^{*-}}\xspace}

\def\Bz      {\ensuremath{B^0}\xspace}
\def\Bzb     {\ensuremath{\Bbar^0}\xspace}
\def\BzBzb   {\ensuremath{\Bz {\kern -0.16em \Bzb}}\xspace}
\def\Bu      {\ensuremath{B^+}\xspace}
\def\Bub     {\ensuremath{B^-}\xspace}

\def\Bpm     {\ensuremath{B^\pm}\xspace}

\def\BpBm    {\ensuremath{\Bu {\kern -0.16em \Bub}}\xspace}

\def\Bbar    {\kern 0.18em\overline{\kern -0.18em B}{}\xspace}

\begin{document}
\vspace*{4cm}
\begin{flushleft}
BABAR-PROC-08/023\\ 
SLAC-PUB-13240\\
\end{flushleft}

\vspace*{2cm}
\title{HADRONIC B DECAYS AT \babar\ AND BELLE}

\author{ VINCENZO LOMBARDO\\ 
\vspace{0.5cm}
(Representing the \babar\ and BELLE Collaborations) }
\vspace*{1.0cm}
\address{Dipartimento di Fisica and INFN Milano, via Celoria 16, Milano (Italy)\\
E-mail: vincenzo.lombardo@mi.infn.it }

\maketitle\abstracts{
We review recent results of the \babar\ and Belle Collaborations on the
the $\alpha$ and $\gamma$ angles of the unitary triangle, on the $B
\rightarrow  K \pi \pi$ Dalitz-plot analyses, and on the searches for
baryonic B decays and for $B \rightarrow D\bar{D}$ decays.}

\section{Introduction}

In this review  we  present recent results  from the \babar\ and Belle
Collaborations on hadronic $B$ decays. The results are organized in 4
sections according to their theoretical interest. In section 2 we describe 
two recent results related to the angle $\alpha$ of the unitarity triangle (UT) 
of the Cabibbo-Kobayashi-Maskawa (CKM) quark mixing matrix~\cite{CKM}, 
in section 3 two updated $B \rightarrow K \pi \pi$ Dalitz-plot analyses, in section 4 several new or updated measurements related to the CKM angle $\gamma$, and in section 5 two searches for baryonic B decays and for $B \rightarrow D^+D^-$ and $\bar{D}^0D^0$ decays.   

\section{Measurements related to the CKM angle $\alpha$} 

Hadronic B decays from $\bar{b} \rightarrow \bar{u}u\bar{d}$ transitions provide the most 
direct information about the weak phase $\alpha$ governing the interference between 
$B^0-\bar{B}^0$ mixing and B decay amplitudes. The difficulty in 
extracting $\alpha$ in these $B$ decay modes is due to the presence of
subleading penguin amplitudes with a different 
weak phase than that of the dominant tree amplitudes. For this  reason the measured 
angle $\alpha_{eff}$, obtained from time-dependent \CP analyses, would differ 
from the UT angle $\alpha$. This difficulty can be overcome by using 
symmetries, either isospin or approximate flavour SU(3)~\cite{GL,Gross,GZ}. In the first 
approach isospin relations can be used to interpret results in terms of a constraint on $\alpha$.
The second approach can be used to extract an upper bound on $|\alpha - \alpha_{eff}|$ 
in terms of branching fractions of a complete set of SU(3) related processes where penguin 
amplitudes are expected to be enhanced.\\ 
The Belle experiment has recently presented at ``Les Rencontres de Physique de la Vallée d'Aoste'' the analysis of neutral B mesons decays to $\rho^0 \rho^0$ using 657 million $\bar{B}B$ events. The measurements of branching fraction and \CP\-violating asymmetries of $B^0\rightarrow \rho^0 \rho^0$ are needed to complete the $B\rightarrow \rho\rho$ isospin relations which allow to constraint the value of $\alpha$ as measured  in $B^0 \rightarrow \rho^+ \rho^-$. The measured upper limit (UL) at 90\% confidence level for the branching fraction of $B^0\rightarrow \rho^0 \rho^0$ is $1.0\times 10^{-6}$. With no  significant signal, this mode was assumed to be longitudinally 
polarized in order to obtain the most conservative UL. This UL is consistent with the \babar\ result 
contributed to the $XIII^{rd}$ International Symposium on Lepton and Photon, 2007~\cite{rhorho2}.\\ 
\babar\ has recently published the first observation for two new
charmless decay
 modes, $B^+ \rightarrow a_1(1260)^+ K^0$ 
and $B^0 \rightarrow a_1(1260)^- K^+$. The branching fractions are
$(34.9\pm 5.0\pm4.4)\times10^{-6}$ and
 $(16.3\pm 2.9\pm2.3)\times 10^{-6}$ with a significance of 6.2 and 5.1
standard deviations ($\sigma$), respectively \cite{a1k}. The
first uncertainty is statistical and the second systematic. Charge asymmetries are found to be consistent with zero. These new results together with the time-dependent \CP\ parameters extracted from  neutral B mesons decays to $a_1(1260) \pi$~\cite{a1pi} and with measurements of the \CP-averaged decay rates $B\to K_1(1270)\pi$ and $B\to K_1(1400)\pi$ will allow to provide a new and independent measurement of the CKM angle $\alpha$~\cite{zupan}.

\section{$B \rightarrow K \pi \pi$ Dalitz-plot analyses}

Two Dalitz-plot analyses of B meson decays to $K\pi\pi$ have been
updated by \babar. These  $B^0 \to K^+ \pi^- \pi^0$ and $B^{\pm} \to
K^{\pm} \pi^{\mp} \pi^{\pm}$ decay modes are interesting for several reasons. 
Two anomalous features measured in $b\rightarrow s$ penguin dominated transitions 
have attracted substantial interest in recent years.  \CP\ asymmetries in these 
modes show a hint of systematic deviation from standard model predictions and 
direct \CP\ asymmetries in $K \pi$ are hard to explain \cite{SysDeviation}. 
Moreover the fact that $K^*\pi$ amplitude can be used to constrain the $(\bar{\rho},\bar{\eta})$ plane of the UT~\cite{ciuchini,gronau} and also the possibility to study the nature of the scalar $f_x(1300)$ make these modes very attractive. 
With a sample of 231.8 million \BB decays the \babar\ Collaboration has measured the magnitudes and phases of the 
intermediate resonant and non-resonant amplitudes for $B^0$ and $\bar
B^0$ decays  and determined the corresponding 
\CP-averaged branching fractions and charge asymmetries \cite{kpipi}. The inclusive
branching fraction and the global \CP-violating charge asymmetry are measured to be 
\calB$(B^0 \to K^+ \pi^- \pi^0) = (35.7^{+2.6}_{-1.5}\pm2.2)\times10^{-6}$ and $\calA_{\CP}=-0.030^{+0.045}_{-0.051}\pm 0.055$, respectively. The branching fraction of the decays $B^0 \to K^{*0}(892)\pi^0$ is observed with a significance of 5.6 $\sigma$, the measured branching fraction of the decay $B^0\to K^{*+}(892)\pi^-$ is in agreement with previous \babar\ and Belle measurements \cite{BaBarPrev,BelleRes}. No evidence for direct \CP-violation in any resonant sub-decay has been found and the decay rate of $B^0\to \rho^- K^+$ is found to be lower by 2 $\sigma$ with respect to the Belle measurement~\cite{BelleRes}.\\
The Dalitz-plot analysis of the  charged B mesons decays to $K^{\pm} \pi^{\mp} \pi^{\pm}$, using a sample of 383 million \BB pairs, has found an evidence above 3 $\sigma$ of direct \CP-violation in $B^+ \to \rho^0 K^+$, with a \CP\ violation parameter $\calA_{\CP}=(+0.44 \pm 0.10 \pm0.04^{+0.05}_{-0.13})$ where the third uncertainty represent the model dependence~\cite{kpipi2}. The total branching fraction \calB $(B^+ \to K^+ \pi^- \pi^+)=(54.4\pm1.1\pm4.5\pm0.7)\times10^{-6}$ is compatible with the Belle measurement \cite{BellePrev}. The $K^{*0}(892)\pi^+$, $(K\pi)^{*0}_0 \pi^+$ and $K^{*0}_2(892) \pi^+$ charge asymmetries are all consistent with zero, as expected \cite{smallCP}. Contributions from $f_2(1270)K^+$ and $f_x(1300)K^+$ are found to be necessary to obtain a good fit to the data with $f_x(1300)$ being a scalar with parameters $m_{f_{x}}=(1479\pm8)\mevcc$ and $\Gamma_{f_x}=(80\pm19)\mev$ where the errors are statistical only.

\section{Measurements related to the CKM angle $\gamma$}

The UT angle $\gamma$  is related to the complex phase of the CKM matrix element $V_{ub}$ through $V_{ub}=|V_{ub}|e^{-i\gamma}$. Over the years various methods have been proposed to measure the angle $\gamma$ using either charged and neural B decays. The GLW method \cite{GLW} has been implemented to work with intermediate $D^0$ and $\bar D^0$ mesons decay to a $\CP$ eigenstate. In the ADS method~\cite{ADS} the $D^0$ from the favored $b \to c$ amplitude is reconstructed in the doubly-Cabibbo suppressed final state, while the $\bar D^0$ from
the $b \to u$ suppressed amplitude is reconstructed in the favored final state. In the GGSZ (Dalitz) method~\cite{GGSZ} the $D^0$ and the $\bar D^0$ are reconstructed in the same $K^0_S \pi^+\pi^-$ three body final state. This method is based on the analysis of the $K^0_S \pi^+\pi^-$ Dalitz distribution and can to some extent be considered as a mixture of the two previous methods. \\
\babar\ has recently reported the first analysis extracting $\gamma$ from neutral B meson decays in $B^0 \to D^0(\bar{D}^0) K^{*0}$ with $D^0 \to K_S \pi^+\pi^-$ \cite{Sordini}. The data sample consists of 371 million \BB events. The charge of the kaon from $K^*$ tags the flavor of the neutral B. The $D^0 (K_S \pi^+\pi^+)$ mesons are analyzed with a well-known Dalitz model plus a K-Matrix formalism for the S-wave component of the $\pi^+ \pi^-$ system. The interference of the signal decay amplitude with other $B^0 \to \bar(D)^0(K\pi)^*_{non-K*}$ amplitudes is taken into account using effective variables. The value of the angle $\gamma$ as a function of $r_S$, the magnitude of the ratio between  $b \to  u$ and $b \to c$ amplitudes have been extracted. Combining this result with the available information on $r_S$, $\gamma$ was found to be $(162\pm56)^\circ$ or $(342\pm56)^\circ$ and $r_S <0.55$ at 95\% probability~\cite{Sordini}.\\ 
An improved measurement of the CKM \CP-violating phase $\gamma$
through a Dalitz-plot analysis of neutral D meson decays to $K_S^0
\pi^+\pi^-$ and $K_S^0 K^+ K^-$ produced in the processes $B^{\mp} \to
D K^{\mp}$, $B^{\mp} \to D^* K^{\mp}$ with $D^* \to D \pi^0, D \gamma$
and $B^{\mp} \to D K^{*\mp}$ with $K^{*\mp} \to
K_S^0\pi^{\mp}$ has recently been updated by the \babar\
Collaboration using a sample of 383 million \BB pairs~\cite{Vidal}. These B meson decays are theoretically clean and are unlikely to be affected by new
physics. The extracted CKM angle $\gamma$ is $ (76 \pm 22 \pm 5 \pm
5)^{\circ}$ (mod $180^{\circ}$) where the first error is statistical,
the second is the experimental systematic uncertainty and the third reflects the uncertainty on the description of the Dalitz-plot distribution. This result has a significance of direct \CP violation ($\gamma \neq 0$ ) of 3.0 $\sigma$. \\
A measurement of the decay $B^{\pm} \to D_{\CP}~K^{\pm}$ has recently
been updated by the \babar\ Collaboration using a sample of 382
million \BB events~\cite{Marchiori}. The $D^0_{\CP}$ mesons have been reconstructed in both a non-\CP\ eigenstate and in \CP (\CP-even and \CP-odd) eigenstates. The ratios ($R_{\CP\pm}$) of charge-averaged partial rates and of the partial-rate of the charge 
asymmetries ($A_{\CP \pm}$) have been measured, $A_{\CP+}=0.27\pm0.09(stat)\pm(0.04)(syst)$, $A_{\CP-}=-0.09\pm0.09(stat)\pm(0.02)(syst)$, $R_{\CP+}= 1.06\pm0.10(stat)\pm0.05(syst)$ and $R_{\CP-}= 1.03\pm0.10(stat)\pm0.05(syst)$.\\ 
\babar\ has reported an improved measurement of the branching
fractions of neutral B meson decays to $D_s^{(*)+} \pi^-$ and
$D_s^{(*)-} K^{+}$ and performed the first branching fraction and
polarization measurement of $D_s^{(*)+} \rho^-$  and $D_s^{(*)-}
K^{*+}$ in a sample of 381 million \BB pairs~\cite{Kol}. The measured branching fractions and longitudinal polarization fractions for the new modes are \calB$(B^0 \to D_s^{(*)+} \rho^-) = (4.4^{+1.3}_{-1.2}\pm0.8)\times 10^{-5}$, $f_L(B^0 \to D_s^{(*)+} \rho^-) = (0.86^{+0.26}_{-0.28}\pm0.15)$, \calB$(B^0 \to D_s^{-} K^{*+}) = (3.6^{+1.0}_{-0.9}\pm0.4)\times 10^{-5}$, \calB$(B^0 \to D_s^{*-} K^{*+}) = (3.0^{+1.4}_{-1.2}\pm0.3)\times 10^{-5}$ and $f_L(B^0 \to D_s^{(*)-} K^{*+}) = (0.96^{+0.38}_{-0.31}\pm0.08)$.
 
\section{Search for baryonic B decays and  $B \rightarrow D^+D^-$ and $\bar{D^0}D^0$}

Several baryonic B decays have been observed over the past years by \babar\ and Belle. A search for the doubly charmed baryonic decays $\bar{B}^0 \to \Lambda_{c}^+ \bar{\Lambda}_{c}^-$ at Belle using a data sample of 520 million \BB events is presented. No significant signal has been found and an upper limit of \calB$(\bar{B}^0 \to \Lambda_c^+ \bar{\Lambda}_c^-) < 6.2\times 10^{-5}$ at 90\% confidence level has been set. The result is significantly below the na\"ive extrapolation from the branching fraction of the decay $B^- \to \Xi_c^0 \bar{\Lambda}_c^-$ assuming a simple factor of $|V_{cd}/V_{cs}|^2$ Cabibbo-suppressed~\cite{lambda}. \\ 
Belle has also reported the branching fraction measurements of $B^0 \to p \bar{p} K^{*0}$ and $B^+ \to p \bar{p} K^{*+}$ with a sample of 535 million $\BB$ pairs. The measured branching fractions are \calB$(B^0 \to p \bar{p} K^{*0}) = (1.18^{+0.29}_{-0.25}(stat.)\pm0.11(syst.))\times10^{-6}$ and \calB$(B^+ \to p \bar{p} K^{*+}) = (3.38^{+0.73}_{-0.60}(stat.)\pm0.39(syst.))\times10^{-6}$ with a statistical significance of 7.2 $\sigma$ and 8.8 $\sigma$, respectively. The polarization of the $K^*$ contains important information on the decay dynamics; the $K^{*0}$ meson is found to be almost $100\%$ in the helicity zero state, compared to  $(32\pm17\pm9)\%$ for the $K^{*+}$ meson \cite{ppbar}. \\ 
An improved measurement of the $B^+ \to D^+ \bar{D^0}$ and $B^0 \to D^0 \bar{D^0}$ decays based on 657 million $\BB$ events is also reported~\cite{DDbar}. The measured branching fraction is \calB$(B^+ \to D^+ \bar{D^0}) = (3.85\pm0.31\pm0.38)\times 10^{-4}$ with charge asymmetry $A_{\CP}(B^+ \to D^+ \bar{D^0}) = 0.00\pm0.08\pm0.02$. An upper limit for $B^0 \to D^0 \bar{D^0}$ has been set for the decay \calB$(B^0 \to \bar{D^0}D^0)<0.42 \times 10^{-4}$.


\begin{thebibliography}{99}
\bibitem{a1}
For the $a_1(1260)$ meson we use the short notation ${a_1}$.
\end{thebibliography}
\end{document}